\RequirePackage{filecontents}

\RequirePackage{snapshot}
\documentclass[aps,reprint,showpacs,secnumarabic,superscriptaddress,footinbib]{preprint}

\def\Preprint{}
\renewcommand\documentclass[2][]{}
\let\origparagraph=\paragraph
\AtBeginDocument{%
  \let\paragraph=\origparagraph
  \let\ack=\acknowledgments
}
\makeatletter
\def\paragraph#1{%
  \@startsection
    {paragraph}%
    {4}%
    {\parindent}%
    {\z@}%
    {-1sp}%
    {\normalfont\normalsize\itshape}{#1.---}%
}%
\makeatother
\documentclass[fleqn]{iopart}

\usepackage[T1]{fontenc}
\expandafter\let\csname equation*\endcsname\relax
\expandafter\let\csname endequation*\endcsname\relax
\usepackage{amsmath}
\allowdisplaybreaks
\usepackage{txfonts}
\usepackage{tgtermes}
\usepackage[altg]{qtxmath}
\renewcommand*{\vec}[1]{\boldsymbol{#1}}
\usepackage{braket}
\usepackage{microtype}
\usepackage{graphicx}
\graphicspath{{figures/}}
\usepackage[breaklinks=true,
pdfauthor={Heiko Bauke, Sven Ahrens, Christoph H. Keitel, and Rainer Grobe},
pdftitle={Electron-spin dynamics induced by photon spins}]{hyperref}
\usepackage{ifthen}
\usepackage{textcomp}
\usepackage[dvipsnames]{color}

\ifdefined\Preprint
\else
\makeatletter
\def\@makefnmark{\hbox{\@textsuperscript{\normalfont\@thefnmark}}}
\def\@makefnmark{\hbox{$^{\fnsymbol{footnote}}\m@th$}}
\let\mathindent\@mathmargin \@mathmargin=5.5pc
\makeatother
\fi

\newcommand*{\I}{\mathrm{i}}

\providecommand*{\e}{\mathrm{e}}
\newcommand*{\trans}[1]{#1^{\mathsf{T}}}
\newcommand*{\E}{\mathcal{E}}

\newcommand*{\acommut}[2]{\left\{#1,#2\right\}}

\begin{document}

\title{Electron-spin dynamics induced by photon spins}

\ifdefined\Preprint

\author{Heiko Bauke}
\email{heiko.bauke@mpi-hd.mpg.de}
\affiliation{Max-Planck-Institut f{\"u}r Kernphysik, Saupfercheckweg~1,
  69117~Heidelberg, Germany} 
\author{Sven Ahrens}
\affiliation{Max-Planck-Institut f{\"u}r Kernphysik, Saupfercheckweg~1,
  69117~Heidelberg, Germany} 
\author{Christoph H. Keitel}
\affiliation{Max-Planck-Institut f{\"u}r Kernphysik, Saupfercheckweg~1,
  69117~Heidelberg, Germany} 
\author{Rainer Grobe}
\affiliation{Max-Planck-Institut f{\"u}r Kernphysik, Saupfercheckweg~1,
  69117~Heidelberg, Germany} 
\affiliation{{Intense Laser Physics Theory Unit and Department of
    Physics, Illinois State University, Normal, IL 61790-4560 USA}} 

\else

\author{Heiko Bauke$^1$,
  Sven Ahrens$^1$,
  Christoph H Keitel$^1$ and
  Rainer Grobe$^{1, 2}$}
\address{$^1$ Max-Planck-Institut f{\"u}r Kernphysik, Saupfercheckweg~1,
  69117~Heidelberg, Germany}
\address{$^2$ Intense Laser Physics Theory Unit and Department of
  Physics, Illinois State University, Normal, IL 61790-4560 USA}
\ead{heiko.bauke@mpi-hd.mpg.de}

\fi

\date{\today}

\begin{abstract}
  Strong rotating magnetic fields may cause a precession of the
  electron's spin around the rotation axis of the magnetic field. The
  superposition of two counterpropagating laser beams with circular
  polarization and opposite helicity features such a rotating magnetic
  field component but also carries spin. The laser's spin density, that
  can be expressed in terms of the laser's electromagnetic fields and
  potentials, couples to the electron's spin via a relativistic
  correction to the Pauli equation. We show that the quantum mechanical
  interaction of the electron's spin with the laser's rotating magnetic
  field and with the laser's spin density counteract each other in such
  a way that a net spin rotation remains with a precession frequency
  that is much smaller than the frequency one would expect from the
  rotating magnetic field alone. In particular, the frequency scales
  differently with the laser's electric field strength depending on if
  relativistic corrections are taken into account or not. Thus, the
  relativistic coupling of the electron's spin to the laser's spin
  density changes the dynamics not only quantitatively but also
  qualitatively as compared to the nonrelativistic theory. The
  electron's spin dynamics is a genuine quantum mechanical relativistic
  effect.

  \vspace{\baselineskip}%
  \vspace{-2.5\baselineskip}%
\end{abstract}


\ifdefined\Preprint
\maketitle
\fi


\section{Introduction}
\label{sec:intro}

Driven by recent developments of novel light sources that envisage to
provide field intensities in excess of $10^{20}\,\mathrm{W/cm^2}$ and
field frequencies in the \mbox{x-ray} domain
\cite{Altarell:etal:2007:XFEL,Yanovsky:Chvykov:etal:2008:Ultra-high_intensity-,McNeil:Thompson:2010:X-ray_free-electron,Emma:Akre:etal:2010:First_lasing,Mourou:Tajima:2011:More_Intense,Mourou:Fisch:etal:2012:Exawatt-Zettawatt_pulse}
relativistic light matter interaction has become an active field of
experimental and theoretical research in recent years
\cite{Brabec:2008:Strong_Field_Laser_Physics,Ehlotzky:Krajewska:Kaminski:2009:Fundamental_processes,Di-Piazza:Muller:etal:2012:Extremely_high-intensity}.
The spin degree of freedom of a bound or free electron may show
nontrivial dynamics or may influence the electron's motion in laser
fields at relativistic intensities.  Spin effects for electrons in
intense linearly polarized laser fields were studied in
\cite{Walser:Urbach:etal:2002:Spin_and,Boca:Dinu:etal:2012:Spin_effects}.
Distinct spin dynamics has been found in the relativistic Rabi effect
\cite{Hammond:2012:Spin_flip} as well as in the relativistic
Kapitza-Dirac effect
\cite{Ahrens:Bauke:etal:2012:Spin_Dynamics,Ahrens:Bauke:etal:2013:Kapitza-Dirac_effect};
in \cite{Skoromnik:Feranchuk:etal:2013:Collapse-and-revival_dynamics}
collapse-and-revival dynamics for the spin evolution of laser-driven
electrons was predicted at $10^{18}\,\mathrm{W/cm^2}$, and also the
dynamics of plasmas can be modified by spin effects
\cite{Brodin:Marklund:etal:2011:Spin_and}.  Furthermore, the spin
orientation relative to the polarization of the driving laser field
affects the ionization rate of highly charged ions
\cite{Klaiber:Yakaboylu:etal:2014:Spin_dynamics}, correlations between
electron spin and photon polarization in bremsstrahlung have been
measured \cite{Tashenov:Back:etal:2011:Measurement_of} and
pair-production rates differ for partices with spin one half and spin
zero
\cite{Cheng:Ware:etal:2009:Pair_creation,Villalba-Chavez:Muller:2013:Photo-production}.

Additionally, the electron also light carries angular momentum
\cite{Jackson:1998:Classical_Electrodynamics}. The (classical) densities
of spin and orbital angular momentum of light can be expressed in terms
of the laser's electromagnetic fields and potentials.  Spin and orbital
angular momentum of light are of particular interest in current research
\cite{Padgett:Allen:2000:twist,Torres:Torner:2011:Twisted_Photons,Andrews:Babiker:2012:Angular_Momentum_of_Light,Fernandez-Corbaton:Zambrana-Puyalto:etal:2012:Helicity_and}
and it appears natural to ask if the light's spin density or orbital
momentum density can couple to massive particles causing observable
effects.  In a pioneering work by Beth
\cite{Beth:1936:Mechanical_Detection} circularly polarized light was
transformed into linear polarization removing spin angular momentum from
the light beam and giving a measurable torque.  Furthermore, in
\cite{He:Friese:etal:1995:Direct_Observation} a coupling of the orbital
angular momentum of light to microscopic (classical) objects causing
rotational or spinning motion was demonstrated and in
\cite{Raeliarijaona:Singh:etal:2013:Predicted_Coupling} a direct
coupling between the angular momentum of light and magnetic moments was
predicted by classical considerations.

In this contribution, we examine electrons and the quantum dynamics of
their spin in a standing light wave formed by two counterpropagating
laser beams with elliptical polarization. We will show that the
electromagnetic wave's spin density couples to the electron's spin
causing electron spin precession.  This precession is identified as a
genuine relativistic effect that can only be explained by treating the
electron quantum mechanically and taking into account relativistic
corrections to the Pauli equation.  The predicted spin effect may be
realized by employing hard x-ray laser pulses of ultra high power with a
stable pulse shape over a few hundred or more cycles.

\section{Elliptically polarized laser beams}
\label{sec:setup}%

We consider an electron in two counterpropagating elliptically polarized
laser beams with the same wavelength $\lambda$ and the same amplitude
$\hat{E}$ but having opposite helicity.  The electric and magnetic field
components of the individual beams are given by
\begin{subequations}
  \label{eq:elliptic_fields}
  \begin{align}
    \label{eq:elliptic_fields_E}
    \vec E_{1,2}(\vec r, t) & = \hat{E} \left(
      \cos(kx\mp \omega t)\vec e_y + \cos(kx\mp \omega t\pm\eta)\vec e_z
    \right) \,,\\
    \label{eq:elliptic_fields_B}
    \vec B_{1,2}(\vec r, t) & = \frac{\hat{E}}{c} \left(
      \mp\cos(kx\mp \omega t\pm\eta)\vec e_y \pm \cos(kx\mp \omega t)\vec e_z
    \right)\,,
  \end{align}
\end{subequations}
with the wave number $k=2\pi/\lambda$, the angular frequency
$\omega=kc$, the speed of light $c$, the position $\vec r
=\trans{(x,y,z)}$, and the time $t$, and $\vec e_x$, $\vec e_y$, and
$\vec e_z$ denoting unit-vectors along the coordinate axes.  The
parameter $\eta\in[0,\pi/2]$ determines the degree of ellipticity with
$\eta=0$ corresponding to linear polarization and $\eta=\pi/2$ to
circular polarization.  In the latter case, the total electric and the
magnetic components $\vec E_{1}(\vec r, t)+\vec E_{2}(\vec r, t)$ and
$\vec B_{1}(\vec r, t)+\vec B_{2}(\vec r, t)$ are parallel to each other
and rotate around the propagation direction.
The laser fields' intensity is independent of the ellipticity and equals
\begin{math}
  I = \varepsilon_0c\hat{E}^2
\end{math}.

The spin angular momentum of light can be associated with its circular
or elliptical polarization. Introducing the Coulomb gauge magnetic
vector potentials $\vec A_{1,2}$ of the
fields~\eqref{eq:elliptic_fields}
\begin{equation}
  \label{eq:vector_fields}
  \vec A_{1,2}(\vec r, t) = -\frac{\hat{E}}{\omega} \left(
    \mp\sin(kx\mp \omega t)\vec e_y \mp \sin(kx\mp \omega t\pm\eta)\vec e_z
  \right) 
\end{equation}
we find that each laser field carries a spin density of
\begin{math}
  \mbox{$\varepsilon_0 \vec E_{1,2}\times \vec A_{1,2}$} =
  {\varepsilon_0 \hat{E}^2\lambda\sin\eta}\,\vec{e}_x/({2\pi c})
\end{math}.  Thus, the spin
density 
of the two plane waves in \eqref{eq:elliptic_fields} points along the
propagation axis of the two waves and is proportional to the sine of the
ellipticity parameter $\eta$. 
Note that the definition of the spin angular momentum of light has been
discussed controversially in the literature
\cite{Mansuripur:2011:Spin_and} and the total photonic spin density is
\begin{equation}
  \label{eq:photonic_spin_dens}
  \varepsilon_0\vec E_{1}\times\vec A_{1} +
  \varepsilon_0\vec E_{2}\times\vec A_{2} =
  \frac{\varepsilon_0 \hat{E}^2\lambda\sin\eta}{\pi c}\vec{e}_x \,,
\end{equation}
which is compatible with a now commonly accepted definition of the
photonic spin density
\cite{Stewart:2005:Angular_momentum,Barnett:2011:On_six,Cameron:Barnett:2012:Electric-magnetic_symmetry,Bliokh:Bekshaev:etal:2013:Dual_electromagnetism}
but it does \emph{not} equal
\begin{equation}
  \varepsilon_0\vec E\times\vec A =
  2\cos^2kx\,
  \frac{\varepsilon_0 \hat{E}^2\lambda\sin\eta}{\pi c}\vec{e}_x \,,
\end{equation}
where $\vec E=\vec E_{1}+\vec E_{2}$ and $\vec A=\vec A_{1}+\vec A_{2}$
denote the total electric field and the total magnetic vector
potential. Nevertheless, the quantity $\varepsilon_0\vec E\times\vec A$
is meaningful for characterizing the electromagnetic field's degree of
polarization because it equals \eqref{eq:photonic_spin_dens} on average
over a wavelength.

\section{Numerical simulations and results}
\label{sec:numerical}

Due to the nonrelativistic electron velocities we assume that the fields
act on the electron but there is no backreaction of the electron to the
laser fields.  This allows us to treat the electromagnetic fields via
the classical vector potential $\vec A(\vec r, t)$.  The evolution of
the electron of mass $m$ and charge $q = -e$ in the laser field,
however, will be described fully quantum mechanically via the Dirac
equation
\begin{equation}
  \label{eq:Dirac_equation}
  \I \hbar\dot \Psi(\vec r, t)  = 
  \left(
    c \vec \alpha\cdot\big(
    -\I\hbar\vec\nabla - q \vec A(\vec r, t)
    \big) + m c^2 \beta
  \right) \Psi(\vec r, t)
\end{equation}
with the Dirac matrices
$\vec\alpha=\trans{(\alpha_x,\alpha_y,\alpha_z)}$ and $\beta$
\cite{Gross:2004:Relativistic_Quantum,Thaller:2005:Advanced_Visual}. The
combined vector potential of the two counterpropagating elliptically 
polarized fields \eqref{eq:elliptic_fields} is given by
\begin{equation}
  \label{eq:A}
  \vec A(\vec r, t) = 
 -\frac{2w(t)\hat{E}}{\omega}\cos kx
  \left( 
    \sin \omega t\,\vec e_y + \sin(\omega t-\eta)\,\vec e_z
  \right)\,,
\end{equation}
where we have deliberately introduced the window function
\begin{equation}
  \label{eq:turn_on_and_off}
  w(t) = 
  \begin{cases}
    \sin^2 \frac{\pi t}{2\Delta T} & \text{if $0\le t\le \Delta T$,} \\
    1 & \text{if $\Delta T\le t\le T-\Delta T$,} \\
    \sin^2 \frac{\pi (T-t)}{2\Delta T} & \text{if $T-\Delta T\le t\le T$,} 
  \end{cases}
\end{equation}
which allows for a smooth turn-on and turn-off of the electromagnetic
fields.  The variables $T$ and $\Delta T$ denote the total interaction
time and the time of turn-on and turn-off.

Taking into account the quasi one-dimensional sinusoidal structure of
the vector potential (\ref{eq:A}) allows us to cast the partial
differential equation (\ref{eq:Dirac_equation}) into a set of coupled
ordinary differential equations
\cite{Ahrens:Bauke:etal:2013:Kapitza-Dirac_effect}.  For this purpose,
we make the ansatz
\begin{equation}
  \label{eq:wave-function}
  \Psi(\vec r, t) = \sum_{n,\gamma} c_n^\gamma(t) \psi_n^\gamma(\vec r)
\end{equation}
with $\gamma\in\{+\uparrow, -\uparrow, +\downarrow, -\downarrow\}$ and the
basis functions
\begin{equation}
  \label{eq:dirac-momentum-eigen}
  \psi_n^{\gamma}(\vec r) = 
  \sqrt{\frac{k}{2 \pi}} \,u_n^\gamma \e^{\I n k x}\,,
\end{equation}
with $u_n^\gamma$ defined as
\begin{subequations}
  \label{eq:dirac-bi-spinors}
  \begin{align}
    u_n^{+\uparrow/+\downarrow} & = \sqrt{\frac{\E_n + m c^2}{2 \E_n}}
    \begin{pmatrix}
      \chi^{\uparrow/\downarrow} \\[0.67ex]
      \frac{n c k \hbar\sigma_x}{\E_n + m c^2} \chi^{\uparrow/\downarrow}
    \end{pmatrix} \,,\\
    u_n^{-\uparrow/-\downarrow} & = \sqrt{\frac{\E_n + m c^2}{2 \E_n}}
    \begin{pmatrix}
      - \frac{n c k \hbar\sigma_x}{\E_n + m c^2} \chi^{\uparrow/\downarrow} \\[0.67ex]
      \chi^{\uparrow/\downarrow}
    \end{pmatrix}\,,
  \end{align}
\end{subequations}
$\chi^\uparrow = \trans{(1, 0)}$ and $\chi^\downarrow = \trans{(0, 1)}$,
the Pauli matrix $\sigma_x$, and the relativistic energy momentum
relation
\begin{math}
  \E_n = \sqrt{(m c^2)^2 + (n c k\hbar)^2}
\end{math}.  The basis functions (\ref{eq:dirac-bi-spinors}) are common
eigenfunctions of the free Dirac Hamiltonian, the canonical momentum
operator, and the Foldy-Wouthuysen spin operator
\cite{Foldy:Wouthuysen:1950:Non-Relativistic_Limit}. Thus,
$\psi_n^{+\uparrow}(\vec r)$ and $\psi_n^{+\downarrow}(\vec r)$ have
positive energy $\E_n$ each and positive and negative spin one half,
respectively, with respect to the $z$~axis.  Introducing the quadruples
$c_n(t)=\trans{(c_n^{+\uparrow}(t),c_n^{+\downarrow}(t),c_n^{-\uparrow}(t),c_n^{-\downarrow}(t))}$
yields from \eqref{eq:Dirac_equation} and \eqref{eq:wave-function} the
Dirac equation in momentum space
\begin{equation}
  \label{eq:dirac-equation-momentum-space}
  \I \hbar\dot c_n(t) = 
  \E_n \beta c_n(t) + \sum_{n'} V_{n,n'}(t) c_{n'}(t) \,,
\end{equation}
with the interaction Hamiltonian $V_{n,n'}(t)$ and its components 
\begin{multline}
  \label{eq:dirac_interaction-hamiltonian}
  V_{n,n'}^{\gamma,\gamma'}(t) = 
  \frac{w(t)q\hat E}{k} \left(
    \delta_{n,n'-1}+\delta_{n,n'+1}
  \right)\times\\
  \left(
    {u_n^{\gamma}}^\dagger \alpha_y u_{n'}^{\gamma'} \sin\omega t +
    {u_n^{\gamma}}^\dagger \alpha_z u_{n'}^{\gamma'} \sin(\omega t - \eta)
  \right)\,.
\end{multline}
The electron is initially at rest with spin aligned to the $z$ axis,
which corresponds to the initial condition $c_0^{{+}\uparrow}(0)=1$ and
$c_n^\gamma(0)=0$ else.
We solve the Dirac equation (\ref{eq:dirac-equation-momentum-space})
numerically \cite{Ahrens:Bauke:etal:2013:Kapitza-Dirac_effect} starting
from this initial condition 
and calculate the spin expectation value
\begin{equation}
  \label{eq:spin_exp}
  s_z(T) = \frac{\hbar}{2}
  \sum_n 
  |c_n^{+\uparrow}(T)|^2 + |c_n^{-\uparrow}(T)|^2 -
  |c_n^{+\downarrow}(T)|^2 - |c_n^{-\downarrow}(T)|^2
\end{equation}
after the field's amplitude dropped to zero at time $t=T$.\footnote{The
  coefficients $c_n^{-\uparrow}(T)$ and $c_n^{-\downarrow}(T)$ are zero
  within numerical errors.}

\begin{figure}
  \centering
  \includegraphics[scale=0.5]{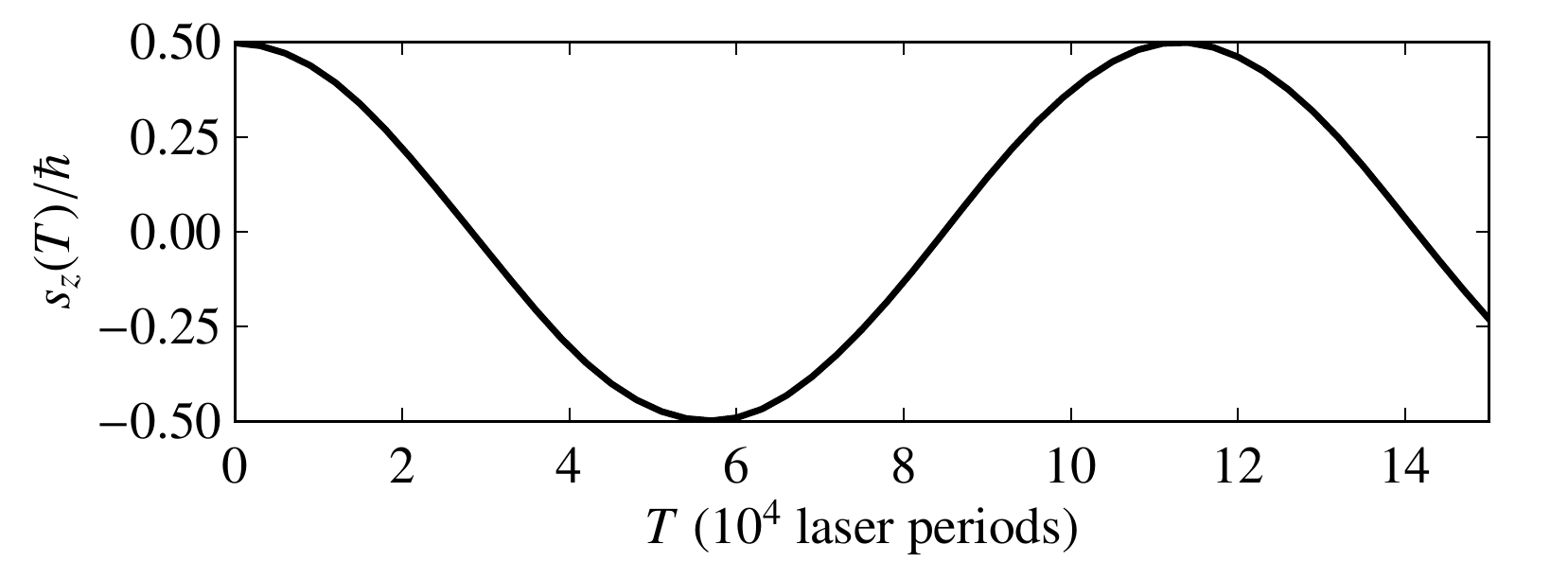}
  \caption{Expectation value of the quantum mechanical electron spin in
    $z$~direction $s_z(T)$ as a function of the total interaction time
    $T$. The wavelength and the amplitude of the applied circularly
    polarized fields is $\lambda=0.992\,\mathrm\AA$ and $\hat
    E=1.38\times10^{14}\,\mathrm{V/m}$, respectively, and the laser's
    switch-on-off interval $\Delta T$ corresponds to 20 laser cycles.}
  \label{fig:spin}
\end{figure}

\begin{figure}
  \centering
  \includegraphics[scale=0.5]{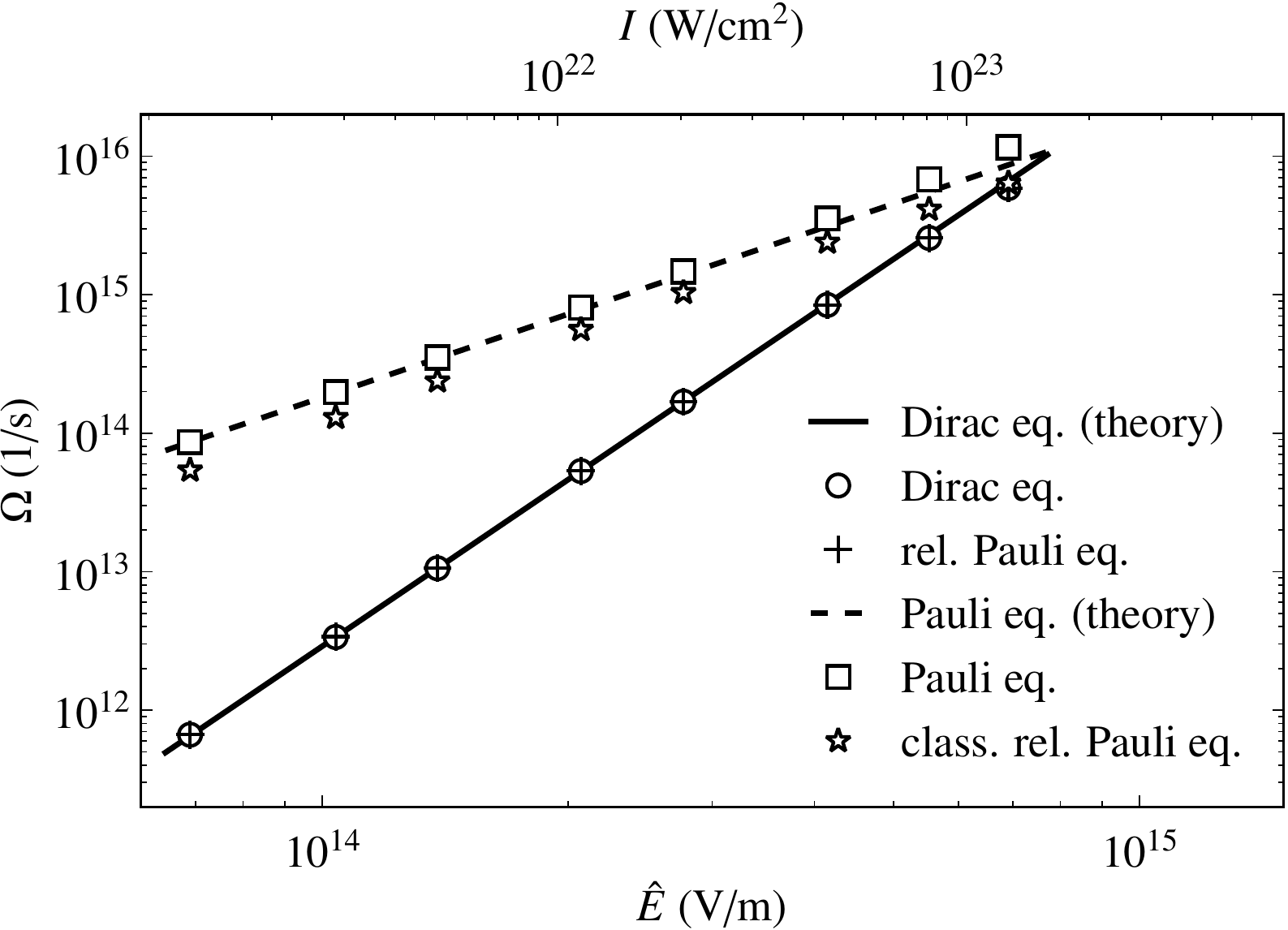}
  \caption{Angular frequency $\Omega$ of the spin precession as a
    function of the laser's electric field strength $\hat{E}$ and its
    intensity $I$ for lasers of the wavelength
    $\lambda=0.992\,\mathrm\AA$.  Depending on the applied
    theory (the Dirac equation \eqref{eq:Dirac_equation}, the
    relativistic Pauli equation \eqref{eq:Pauli_FW_equation}, the
    nonrelativistic Pauli equation, or the classical equation
    \eqref{eq:Pauli_FW_equation_classical}) the spin of an electron in
    two counterpropagating circularly polarized light waves scales with
    the second or fourth power of $\hat{E}$.  Theoretical predictions
    for the spin-precession frequency are given by
    \eqref{eq:Omega_Dirac} and \eqref{eq:Omega_P}, respectively.}
  \label{fig:scaling}
\end{figure}

Let us consider the special case of circular polarization first.  As
shown in Fig.~\ref{fig:spin} the expectation value $s_z(T)$ oscillates
periodically with a period much larger than the laser's period, i.\,e.,
it does not just follow the field adiabatically.  The spin's expectation
value in the $y$~direction oscillates in a similar fashion but with a
$\pi/2$ phase shift.  Thus, the electron's spin precesses around the
$x$~axis, i.\,e., the propagation direction of the laser field.
Considering the interaction Hamiltonian $V_{n,n'} (t)$ as a perturbation
to the free Dirac Hamiltonian, time-dependent perturbation theory
\cite{Ahrens:Bauke:etal:2013:Kapitza-Dirac_effect} for the Dirac
equation \eqref{eq:dirac-equation-momentum-space} yields the angular
frequency $\Omega$ of the spin precession for the case of circular
polarization
\begin{equation}
  \label{eq:Omega_Dirac}
  \Omega = \frac{(q\hat{E})^4\lambda^5}{(2\pi)^5\hbar^2m^2c^5}\,.
\end{equation}
One can show that the application of perturbation theory is justified
provided that both inequalities ${|q|\hat E}<{k^2\hbar c}$ and ${|q|\hat
  E}<{2k m c^2}$ hold.  A comparison of this prediction for the
spin-precession angular frequency with results from the numerical
solution of the Dirac equation \eqref{eq:dirac-equation-momentum-space}
shows a good agreement between theory and simulations, see black solid
line and circles in Fig.~\ref{fig:scaling}.  In particular, the
numerical results confirm that the spin-precession angular frequency
$\Omega$ growths with the \emph{fourth} power of the electric field
strength $\hat E$.

\begin{figure}
  \centering
  \includegraphics[scale=0.5]{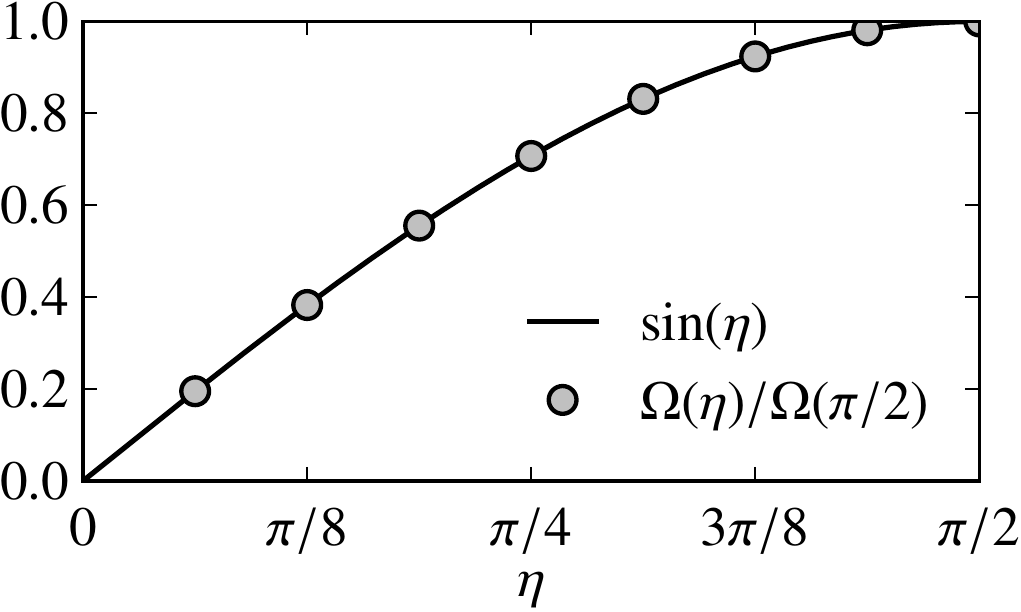}
  \caption{Spin-precession angular frequency $\Omega$ (normalized to the
    precession frequency for circular light) as a function of the
    ellipticity parameter $\eta$.  The electric field strength was set
    to $\hat E=5.53\times10^{14}\,\mathrm{V/m}$; other parameters as in
    Fig.~\ref{fig:spin}.}
  \label{fig:scaling_elliptic}
\end{figure}

While Figs.~\ref{fig:spin} and \ref{fig:scaling} show results for
circular polarization Fig.~\ref{fig:scaling_elliptic} demonstrates the
effect of an ellipticity parameter $\eta\ne\pi/2$ by showing the
spin-precession angular frequency $\Omega$ as a function of $\eta$.  Our
numerical results indicate that $\Omega$ is proportional to $\sin\eta$.
Thus, the spin-precession angular frequency for electrons in
counterpropagating elliptically polarized laser fields is proportional
to the product of the laser field's intensity 
and the spin density 
of the electromagnetic wave.

\section{Electron dynamics in the weakly relativistic limit} 

The Dirac equation \eqref{eq:Dirac_equation} describes the fully
relativistic quantum dynamics of the electron.  It is, however, not very
transparent when it comes to physical interpretation.  It is a standard
procedure to consider a nonrelativistic expansion of the Dirac equation
via a Foldy-Wouthuysen transformation
\cite{Foldy:Wouthuysen:1950:Non-Relativistic_Limit,Frohlich:Studer:1993:Gauge_invariance}.
The resulting expansion
\begin{multline}
  \label{eq:Dirac_FW_equation}
  \I \hbar\dot \Psi(\vec r, t)  = 
  \Bigg( 
  \frac{( -\I\hbar\vec\nabla - q \vec A(\vec r, t) )^2}{2m} -
  \frac{q\hbar}{2m}\vec\sigma\cdot \vec B(\vec r, t) 
  +
  q\phi(\vec r, t) \\ 
  -
  \frac{( -\I\hbar\vec\nabla - q \vec A(\vec r, t) )^4}{8m^3c^2} -
  \frac{q^2\hbar^2}{8m^3c^4}(c^2\vec B(\vec r, t)^2-\vec E(\vec r, t)^2) \\+
  \frac{q\hbar}{8m^3c^2}\acommut{\vec\sigma\cdot\vec B(\vec r, t)}{( -\I\hbar\vec\nabla - q \vec A(\vec r, t) )^2}
  \\
  -
  \frac{q\hbar}{4m^2c^2 }\vec\sigma\cdot \big(\vec E(\vec r, t) \times (
  -\I\hbar\vec\nabla - q\vec A(\vec r, t) )\big)  -
  \frac{q\hbar^2}{8m^2c^2}\vec\nabla\cdot\vec{E}(\vec r, t) 
  \Bigg)\Psi(\vec r, t) \,.
\end{multline}
with $\vec A$ given by \eqref{eq:A}, $\vec B=\vec\nabla\times\vec A$, $\vec
E=-\dot{\vec A}$, and the Pauli matrices
$\vec\sigma=\trans{(\sigma_x,\sigma_y,\sigma_z)}$ allows for a manifest
interpretation of the individual corrections to the nonrelativistic
Pauli equation.  The relativistic corrections $\sim{(-\I\hbar\vec\nabla - q
  \vec A)^4}$ and $\sim\vec\nabla\cdot\vec{E}$ (the so-called Darwin term)
account for relativistic mass effects and field inhomogeneities,
respectively, and the contribution $\sim\vec\sigma\cdot (\vec E \times
(\I\hbar\vec\nabla))$ leads to the well-known spin-orbit effect, while
$\vec\sigma\cdot (\vec E \times \vec A)$ couples the photonic spin
density to the electronic spin.  

Conjecturing that the electromagnetic field's spin density is the only
source for relativistic effects in the electron's spin dynamics, we
consider now the relativistic Pauli equation where only the relativistic
correction due to the field's spin density is included, i.\,e.,
\begin{multline}
  \label{eq:Pauli_FW_equation}
  \I \hbar\dot \Psi(\vec r, t)  = 
  \Bigg(
    \frac{1}{2m}( -\I\hbar\vec\nabla - q \vec A(\vec r, t) )^2 -
    \frac{q\hbar}{2m}\vec\sigma\cdot \vec B(\vec r, t) \\ +
    \frac{q^2\hbar}{4m^2c^2 }\vec\sigma\cdot (\vec E(\vec r, t) \times \vec A(\vec r, t) )
  \Bigg)\Psi(\vec r, t) \,.
\end{multline}
This equation can be cast into a set of ordinary differential equations
by a plane wave ansatz
\cite{Ahrens:Bauke:etal:2013:Kapitza-Dirac_effect} similar to
\eqref{eq:wave-function} for the Dirac equation.  The angular frequency
of the spin precession that results from a numerical solution of
\eqref{eq:Pauli_FW_equation} agrees with the frequency from the full
Dirac equation \eqref{eq:Dirac_equation}; see Fig.~\ref{fig:scaling}.
This justifies our conjecture to utilize \eqref{eq:Pauli_FW_equation} as
an effective equation for the full Dirac equation
\eqref{eq:Dirac_equation} and highlights the role of the electromagnetic
field's spin density.  Furthermore, the scaling of the spin-precession
angular frequency changes qualitatively if the relativistic correction
term in \eqref{eq:Pauli_FW_equation}
is neglected.  In fact, the spin-precession angular
frequency scales with the \emph{second} power of the electric field
strength in this case; see squares in Fig.~\ref{fig:scaling}.

\section{Classical electron-spin dynamics}

The spin precession in the oscillatory field of the two
counterpropagating elliptical laser fields may one remind on the Rabi
effect \cite{Rabi:1937:Space_Quantization}.  This problem can be well
understood by a classical model
\cite{Allen:Eberly:1987:Optical_Resonance}.  In the following we will
adopt this approach to study spin precession in elliptical laser fields.
For simplicity we assume circular polarization ($\eta=\pi/2$).  The
quantum state is delocalized over several laser wavelengths; thus we
mimic the quantum wave packet by an ensemble of classical particles with
spin angular momentum. These particles are placed along the $x$ axis.
The dynamics of a classical electron spin $\vec s$ at fixed position
$\vec r$ in the magnetic field $\vec B(\vec r, t)=\vec B_{1}(\vec r,
t)+\vec B_{2}(\vec r, t)$ is governed by
\begin{equation}
  \label{eq:s_B}
  \dot{\vec s}(t) = \frac{q}{m} \vec s(t)\times\vec B(\vec r, t)\,.
\end{equation}
This equation of motion can be justified via the Pauli equation, i.\,e.,
Eq.~\eqref{eq:Pauli_FW_equation} neglecting the correction term that is
proportional to \mbox{$\vec E\times\vec A$}, by calculating the quantum
mechanical equation of motion for the spin observable in the Heisenberg
picture \cite{Silenko:2008:Foldy-Wouthyusen_transformation}.  For
parameters such that $\Omega_\mathrm{L}=q\hat E/(mc)\ll\omega$, where
$\Omega_\mathrm{L}$ denotes the oscillation frequency of a spin in a
constant magnetic field with amplitude $\hat E/c$, the spin $\vec s(t)$
rotates with a position-dependent angular frequency around the axis of
rotation of the magnetic field.  For the initial condition $\vec s
(0)=\trans{(0,0,\hbar/2)}$ the solution of \eqref{eq:s_B} is given by
\begin{equation}
  \label{eq:s_B_sol}
  {\vec s}(t)= \frac{\hbar}{2}
  \trans{\begin{pmatrix}
    0 ,
    \sin (2\Omega_\mathrm{P}t\sin^2 kx) ,
    \cos (2\Omega_\mathrm{P}t\sin^2 kx) 
  \end{pmatrix}}
\end{equation}
with the angular frequency averaged over a wavelength
\begin{equation}
  \label{eq:Omega_P}
  \Omega_\mathrm{P}=\frac{(q\hat E)^2\lambda}{2\pi m^2c^3}\,.
\end{equation}
The frequency \eqref{eq:Omega_P} agrees with the spin-precession angular
frequency for the nonrelativistic Pauli equation that can be calculated
via time-dependent perturbation theory and that is also confirmed by our
numerical simulations; see squares in Fig.~\ref{fig:scaling}.
Similarly, the relativistic correction in \eqref{eq:Pauli_FW_equation}
to the Pauli equation leads to the classical equation
\begin{equation}
  \label{eq:s_ExA}
  \dot{\vec s}(t) = -\frac{q^2}{2m^2c^2} 
  \vec s(t)\times(\vec E(\vec r, t)\times\vec A(\vec r, t))\,.
\end{equation}
For the initial condition ${\vec s}(0)=\trans{(0,0,\hbar/2)}$ the
solution of \eqref{eq:s_ExA} is given by
\begin{equation}
  \label{eq:s_ExA_sol}
  {\vec s}(t)= \frac{\hbar}{2}
  \trans{\begin{pmatrix}
    0 ,
    \sin (-2\Omega_\mathrm{P}t\cos^2 kx) ,
    \cos (-2\Omega_\mathrm{P}t\cos^2 kx) 
  \end{pmatrix}}\,.
\end{equation}

Note that due to the fast rotation of the magnetic field 
the net effect of the spin term proportional to $\vec B$ in
\eqref{eq:Pauli_FW_equation} is of the same order as the relativistic
correction proportional to $\vec E\times\vec A$.  As a consequence of
\eqref{eq:s_B_sol} and \eqref{eq:s_ExA_sol} a classical spin under the
effect of both spin terms rotates in a short time interval $\Delta t$
around the angle $2\Omega_\mathrm{P}(\sin^2 kx - \cos^2 kx)\Delta t$.
Averaged over a laser wavelength this rotation angle vanishes and the
effects of the two spin terms chancel each other in our classical model.
Thus, the classical model explains how the effect of the laser fields'
spin density $\sim\vec E\times\vec A$ leads to a breakdown of the
quadratic scaling of the spin-precession angular frequency in $\hat E$
that results if only the magnetic field is taken into account.  The
model, however, is not able to reproduce the quartic scaling in $\hat E$
that results from the fully relativistic quantum mechanical Dirac
equation \eqref{eq:Dirac_equation}.  The quartic scaling results as a
genuine quantum effect from the fast temporal oscillations combined with
the spatial modulation of the electromagnetic fields.  Unlike the Rabi
effect it does not solely result from the temporal oscillations of the
electromagnetic fields.

In order to confirm that the precession frequency \eqref{eq:Omega_Dirac}
is of quantum mechanical origin we also solved the coupled classical
equations of motion for the spin $\vec s$, the position $\vec r$, and
the canonical momentum $\vec p$ that correspond to the quantum
mechanical Hamiltonian \eqref{eq:Pauli_FW_equation}, namely
\begin{subequations}
  \label{eq:Pauli_FW_equation_classical}
  \begin{align}
    \dot{\vec r}(t) & = \frac{\partial H(\vec r, \vec p)}{\partial \vec p} \,,\\
    \dot{\vec p}(t) & = -\frac{\partial H(\vec r, \vec p)}{\partial \vec r} \,,\\
    \dot{\vec s}(t) & = \frac{q}{m} \vec s(t)\times\vec B(\vec r, t)
    -\frac{q^2}{2m^2c^2} \vec s(t)\times(\vec E(\vec r, t)\times\vec
    A(\vec r, t)) \,, 
  \end{align}
  with
\ifdefined\Preprint
  \begin{multline}
    H(\vec r, \vec p) = 
    \Bigg(
    \frac{1}{2m}(\vec p - q \vec A(\vec r, t) )^2 -
    \frac{q}{m}\vec s(t)\cdot \vec B(\vec r, t) \\
    +
    \frac{q^2}{2m^2c^2 }\vec s(t)\cdot (\vec E(\vec r, t) \times \vec A(\vec r, t) )
    \Bigg)\,,
  \end{multline}
\else
  \begin{equation}
    H(\vec r, \vec p) = 
    \Bigg(
    \frac{1}{2m}(\vec p - q \vec A(\vec r, t) )^2 -
    \frac{q}{m}\vec s(t)\cdot \vec B(\vec r, t) 
    +
    \frac{q^2}{2m^2c^2 }\vec s(t)\cdot (\vec E(\vec r, t) \times \vec A(\vec r, t) )
    \Bigg)\,,
  \end{equation}
\fi
\end{subequations}
for the initial conditions ${\vec r}(0)=\trans{(0,0,0)}$, ${\vec
  p}(0)=\trans{(0,0,0)}$, and ${\vec s}(0)=\trans{(0,0,\hbar/2)}$.  As
we see in Fig.~\ref{fig:scaling} the spin-precession frequency (stars)
is reduced compared to \eqref{eq:Omega_P} (dashed line) but also scales
quadratically with the electric field amplitude.\footnote{The
  spin-precession frequency scales quadratically with the electric field
  amplitude even if higher relativistic corrections are taken into
  account via a classical equation given in
  \cite{Silenko:2008:Foldy-Wouthyusen_transformation}.}  The action of
the magnetic field on the electron's spin is \emph{not} canceled by the
action due to the laser fields' spin-density because the time that the
point-like electron spends in different regions of the laser field is
not distributed uniformly over a laser wavelength.

\section{Experimental realization}

Let us shortly estimate for what kind of laser parameters the electron's
spin precession in elliptical fields may be observed.  An experimental
realization may utilize intense photon beams at nearfuture x-ray laser
facilities to form standing waves.  Intensity, frequency, and pulse
length have to be carefully adjusted to make the spin precession
detectable.  For a given wavelength $\lambda$ and pulse length $n$
(number of cycles) the required electric field strength is bounded as
\begin{equation}
  \label{eq:bounds_E}
  \left({\frac{(2\pi)^6 c^6\hbar^2m^2}{2nq^4\lambda^6}}\right)^{1/4} < 
  \hat E < 
  \frac{(2\pi)^2c\hbar}{|q|\lambda^2}\,.
\end{equation}
In an experimental test of the predicted effect one has to detect a
change of the electron's spin orientation, which must be large enough to
be measurable. The lower bound in \eqref{eq:bounds_E} results from the
fact that at lower intensities the electron must be confined in the
focus of the counterpropagating laser pulses for more cycles to realize
a full spin flip.  Note that the lower bound in \eqref{eq:bounds_E} can
be made arbitrarily small by inceasing the number of cycles $n$ and/or
raising the wavelength $\lambda$.  Thus, the predicted coupling between
electronic and photonic spins and the resulting electron spin precession
are not intrinsic strong-field effects.  Note that an increase of the
wavelength may has to be accompanied by an extension of the pulse length
$n$, otherwise the lower limit in \eqref{eq:bounds_E} may exceed the
upper limit. To give a specific numerical example, for $n=5000$ the two
bounds in \eqref{eq:bounds_E} coincide for a wavelength of
$\lambda=0.24\,\mathrm{nm}$ yielding $\hat
E=1.32\times10^{14}\,\mathrm{V/m}$ and an intensity of
$I=4.64\times10^{21}\,\mathrm{W/cm^2}$.  If the upper bound in
\eqref{eq:bounds_E} is violated the spin precession becomes anharmonic.

\section{Conclusions} 

We investigated electron motion in two counterpropagating elliptically
polarized laser waves of opposite helicity and found spin precession
around the fields' propagation direction.  The spin precession can be
properly described via the fully relativistic Dirac equation or the
nonrelativistic Pauli equation plus a relativistic correction
proportional to $\vec E\times\vec A$.  Because this quantity is closely
related to the spin angular momentum of elliptically polarized light the
relativistic correction can be interpreted as a coupling of the laser
field's spin density to the electron spin.  The spin-precession
frequency is proportional to the product of the laser field's spin
density and its intensity. The electron-spin dynamics has the remarkable
feature that it is a genuine relativistic quantum effect. The quartic
scaling of the spin-precession frequency with the electric field
strength can not be found within the nonrelativistic Pauli theory; and
also classical models that treat electrons as point-like particles fail
to reproduce the correct spin-precession frequency even if relativistic
effects due to the laser field's spin density are taken into account.
This distinguishes the predicted spin effect from relativistic spin
effects in other setups, e.\,g., in the Kapitza-Dirac effect
\cite{Ahrens:Bauke:etal:2012:Spin_Dynamics,Ahrens:Bauke:etal:2013:Kapitza-Dirac_effect},
where relativistic signatures are induced via relativistic system
parameters, e.\,g., initial electron momenta.

We demonstrated that the coupling of the photonic spin to the electron's
spin is closely related to the well-known spin-orbit effect.  Both are
examples for the interaction of two different kinds of angular momentum,
a generic mechanism that can be observed also in very different
contexts.  For example, in \cite{Zhang:Niu:2014:Angular_Momentum} the
interaction of the angular momentum of phonons in a magnetic crystal
with the angular momentum of the crystal was recently considered and in
\cite{Mondal:Deb:etal:2014:Angular_momentum} it is investigated how
the orbital angular momentum of light affects the internal degrees of
motion in molecules.  In our setup of counterpropagating plane waves,
the spin angular momentum is the electromagnetic field's only angular
momentum.  More realistic focused beams may also carry orbital angular
momentum, which may also couple to the electronic spin and in this way
modify the spin dynamics.  Because of its fundamental nature we expect
that the demonstrated coupling of the photonic spin to the electron's
spin is of interest also for researchers beyond the strong field
community.

\ack We have enjoyed helpful discussions with Prof.~C.~M{\"u}ller.  R.~G.\
acknowledges the nice hospitality during his visit in Heidelberg.  This
work was supported by the NSF.

\ifdefined\Preprint
\else
\pagebreak[1]
\enlargethispage{2.2ex}
\section*{References}
\fi

\bibliography{electron_spin_in_circular_light}

\end{document}